\documentclass[runningheads]{llncs}
\usepackage[T1]{fontenc}
\usepackage{graphicx}
\usepackage{epstopdf}
\usepackage{amsmath,amssymb}
\usepackage{booktabs}
\usepackage{multirow}
\usepackage{multicol}
\usepackage{subcaption}
\usepackage[table]{xcolor}
\usepackage{adjustbox}
\usepackage{url}
\usepackage[hidelinks]{hyperref}
\begin{document}
\title{You Can't Spot a Deepfake---And Neither Can Your Brain Nor Eyes: A Neurophysiological Framework for Deepfake Exploitation of Cognitive Engagement and Implicit Visual Evaluation}
\titlerunning{You Can't Spot a Deepfake---And Neither Can Your Brain Nor Eyes}
%% ---- Set \anonfalse for the CAMERA-READY build, \anontrue for the SHEPHERD build ----
\newif\ifanon
\anonfalse
\ifanon
  \author{Anonymous Author(s)}
  \authorrunning{Anonymous Submission}
  \institute{Paper submitted for double-blind review}
\else
    \author{Cagri Arisoy\inst{1}\thanks{Work performed while a graduate student at Texas A\&M University.} \and Md Imanul Huq\inst{2}\thanks{Corresponding author.} \and Amy W. Hays\inst{2} \and Nitesh Saxena\inst{2}}
    
  %\author{Cagri Arisoy\inst{1} \and Md Imanul Huq\inst{2}\thanks{Corresponding author.} \and Amy W. Hays\inst{2} \and Nitesh Saxena\inst{2}}
  \authorrunning{C. Arisoy et al.}
  \institute{Yozgat Bozok University, 66100 Yozgat, T\"urkiye\\ \email{cagri.arisoy@bozok.edu.tr} \and Texas A\&M University, College Station, TX 77843, USA\\ \email{\{imanulhuq, amy.hays, nsaxena\}@tamu.edu}}
\fi
\maketitle
\thispagestyle{plain}
\begin{abstract}
Deepfakes have rapidly emerged as one of the most pressing threats to information integrity and security, precisely because they are designed to exploit human trust in visual and auditory perception. Yet, little is known about whether humans and their underlying (sub)conscious neuro-physiological processes — can reliably distinguish deepfake vs. real videos.

We introduce \textbf{DECEIVE} (Deepfake Exploitation of Cognitive Engagement and Implicit Visual Evaluation), a framework that models how deepfake videos are validated as adversarial payloads through behavioral and neuro-physiological screening of viewers, and how an attack is iteratively refined by selecting the payloads that evade detection. The framework is dataset agnostic and applies to any synthetic or real media. Such a framework is inherently dual-use: an adversary with equivalent measurements could iterate on candidate manipulations and keep those that evade human detection, which is precisely why the evaluation should be conducted openly and defensively. Measuring which deepfakes already defeat human perception establishes a realistic bound on attacker capability, against which detection tooling, provenance and watermarking mechanisms, and user-facing protections can be assessed.

As an instantiation, we conducted an EEG and eye-tracking study in which participants viewed real, deepfake, and look-alike videos drawn from Celeb-DF and a curated celebrity set, while their behavioral judgments and implicit responses were recorded. Contrary to expectations of subconscious differentiation suggested by prior work on paintings and phishing websites, no statistically significant neuro-physiological differences emerged between real and deepfake videos, although clear distinctions were observed for look-alike videos. Behaviorally, participants accepted 26.68\% of manipulated clips as authentic, rising to 31.94\% for familiar identities, confirming the studied deepfakes as effective adversarial payloads within the DECEIVE framework.

\end{abstract}

\section{Introduction}
\label{sec:intro}
The proliferation of the internet and social media use has led to an unprecedented surge in content generation, with over 1 trillion megabytes of data produced daily by 5 billion users—approximately 66\% of the global population \cite{StatistaResearchDepartment2021}. This trend, accelerated by technological advancements and the COVID-19 pandemic \cite{Perrin2020,roessler2019faceforensics}, has facilitated not only new opportunities for communication but also new challenges in security and privacy. One of the most pressing concerns is deepfake technology, which uses advanced generative techniques such as generative adversarial networks (GANs) and convolutional neural networks (CNNs) to create highly realistic yet synthetic videos. These manipulated media are increasingly used beyond entertainment, with implications for political misinformation, identity theft, fake advertisements, adult content, and privacy violations \cite{northeastern-deepfakes-fake-news,analyticsinsight-deepfake-concerns,theatlantic-deepfake-porn,cameraforensics-deepfakes-privacy}. 

The danger of deepfakes lies in their design: they exploit the very mechanisms of human trust in visual perception, leaving users uncertain about what is authentic and what is manipulated. The escalating prevalence of such videos demands systematic methods to understand and model their impact on human cognition. \textit{How can we effectively determine whether deepfakes will actually work to deceive the users?} Evaluating deepfakes in an attack context allows us to measure their operational effectiveness against human perceptual defenses, providing actionable insight for both threat modeling and countermeasure design.

\smallskip
\noindent \textbf{A First Neuro-physiological Framework to Iteratively Assess Deepfake Susceptibility:} To address this question, we introduce DECEIVE (Deepfake Exploitation of Cognitive Engagement and Implicit Visual Evaluation), which conceptualizes deepfakes as adversarial payloads tested iteratively against human observers at two levels: behavioral compromise (misclassifying real vs.\ fake) and subconscious evasion (absence of distinct EEG or ocular markers). Payloads humans fail to detect at both levels are deemed successful, and the resulting profile of which manipulations bypass human defenses informs threat modeling and countermeasure design. Our expectation of neuro-physiological differences is grounded in predictive-coding and authenticity-judgment research, in which the brain compares sensory input against learned patterns of realism \cite{walsh2020predictive}; examining whether such cues persist for synthetic media probes how deeply deepfakes bypass subconscious perception.

\smallskip
\noindent \textbf{An EEG and Eye-tracking Study to Instantiate DECEIVE against Real-World Deepfakes:} As the first instantiation of the DECEIVE framework, we conducted a controlled laboratory experiment that exposed participants to authentic, deepfake, and look-alike videos while measuring both behavioral judgments and neuro-physiological responses via EEG and eye-tracking. 
In our study, state-of-the-art publicly available deepfake datasets such as YouTube-based Celeb-DF, extended with additional celebrity content, serve as the tested attack vectors. Attack success was defined as behavioral compromise (false acceptance of manipulated videos) and subconscious evasion (absence of distinct neurological or ocular markers differentiating real from fake). Contrary to expectations of subconscious sensitivity, our findings reveal no significant neuro-physiological differences between real and deepfake videos, while look-alike videos did elicit distinct processing. This outcome confirms that high-quality publicly-available deepfakes function as successful adversarial payloads, bypassing both conscious and subconscious defenses.

Cognitive neuroscience shows that the brain processes visual stimuli in complex, category-specific ways \cite{haxby2001,KEIGHTLEY2003585,VUILLEUMIER2005585}. Our study hypothesizes that real, deepfake, and impersonated videos elicit distinct neural and attentional responses. Impersonations may trigger familiarity recognition and detection of subtle inconsistencies, while deepfakes—artificial identity manipulations—could evoke sensitivity to distortions or unnatural features.

\smallskip
\noindent \textbf{Differentiation from Related Prior Work:} Prior studies suggest possible subconscious neural differentiation. Neupane et al. \cite{neupane2014neural,neupane2015multimodal,neupane2017neural} found EEG-based differences in responses to real vs.\ fake phishing websites, while Huang et al. \cite{huang2011human} reported distinct neural activations for authentic versus forged Rembrandt paintings, even without behavioral discrimination. Yet, findings are mixed: Neupane et al. \cite{neupane2019crux} observed no neural differences between real and fake voices, implying some manipulations may escape subconscious detection.

Our study extends prior research using neuro-physiological measures to study real/fake attack susceptibility \cite{neupane2015multimodal,gupta2020eyes,neupane2019crux,arisoy2022human}, focusing on emerging deepfake videos. We examine behavioral and neural responses to real, deepfake, and lookalike (ALIKE) videos using EEG and eye tracking, a combination valued for managing eye-movement artifacts \cite{ploch2012}. The dataset includes familiar (FAM), briefly familiar (BFAM), and lookalike individuals (details in Section~\ref{dataset}). Participants viewed soundless videos without time limits, ensuring attention to visual cues while avoiding auditory confounds. This stepwise focus on visual processing establishes a foundation before multi-modal research. Related work on deepfake voices shows distinct neural mechanisms \cite{Roswandowitz2024}; likewise, we study brain and cognitive processes underlying visual deepfake perception, particularly how they influence cognitive load and attention. Our results show no clear neuro-cognitive or ocular distinction between real and deepfake videos, but marked differences emerge when both are contrasted with lookalike videos. Lookalike/body-double conspiracies are also reported\footnote{https://www.bbc.com/news/uk-68609361}, yet our findings suggest users are less susceptible to them compared to deepfakes. This pattern differs from real–fake website detection studies \cite{neupane2014neural,neupane2015multimodal,neupane2017neural} yet aligns with findings on real–fake voice detection \cite{neupane2019crux}. Table \ref{tab:real-fake} presents a summary outlining our study's findings in comparison to other research endeavors focused on real-fake detection.
 
In our adversarial evaluation, this pattern of results indicates a successful attack: deepfakes bypassed both behavioral and neuro-ocular detection, while look-alike videos were more easily differentiated. Our study explores how the brain’s automatic perception processes respond to synthetic media and whether deepfakes can bypass these subconscious cues that usually help viewers sense authenticity.

\begin{table}[t]
\centering
\caption{Real–fake comparisons across modalities.}
\label{tab:real-fake}
\renewcommand{\arraystretch}{0.9}
\setlength{\tabcolsep}{3pt}
\scriptsize
\begin{tabular}{|p{0.45\columnwidth}|p{0.22\columnwidth}|p{0.22\columnwidth}|}
\hline
\textbf{Artifact type} & \textbf{Neural activation} & \textbf{Behavioral response} \\ \hline \hline
Websites under phishing \cite{neupane2014neural,neupane2015multimodal,neupane2017neural} 
& Present & Nearly absent \\ \hline
Rembrandt paintings \cite{huang2011human} 
& Present & Nearly absent \\ \hline
Fake voices \cite{neupane2019crux} 
& Absent & Nearly absent \\ \hline
\textbf{Deepfake videos (this work)} 
& \textbf{Absent} & \textbf{Partially present} \\ \hline
\end{tabular}
\end{table}

\smallskip
\noindent \textbf{Our Contributions:} Our work shifts focus from the behavioral difficulty of detecting deepfakes to the neuro-physiological mechanisms underlying it. The key finding is that real and deepfake videos elicit no clear neuro-physiological or ocular differences---unlike paintings or websites in prior work---consistent with our behavioral results and with Neupane et al.\ \cite{neupane2019crux} on fake voices, underscoring human vulnerability and the need for automated detection and user training. Our framework can be used to assess the resilience of human viewers against any deepfake set, serving as both an evaluation and a defense tool.

\section{Background Information}
\label{sec:background}
In our study we use several notations to describe the deepfake videos, participants' responses, and eye-tracking measures. All constructs used in this paper are defined at first use in the main text and are collected for reference in Table~\ref{tab:notations}.

\begin{table}[ht]
\footnotesize
\centering
\caption{Summary of notations used throughout the paper.}
\label{tab:notations}
\begin{tabular}{|l|l|}
\hline
\textbf{Notation} & \textbf{Description} \\ \hline\hline
FAM & Familiar videos (widely recognizable public figures) \\ \hline
BFAM & Briefly familiar videos (actors introduced anonymously) \\ \hline
ALIKE & Look-alike videos (genuine doppelg\"angers, unmanipulated) \\ \hline
Celeb-DF & Celeb DeepFake dataset \\ \hline
COT-DB & Curated YouTube celebrity database (this work) \\ \hline
D-Fake & Difficult fake videos (below the accuracy threshold) \\ \hline
E-Fake & Easy fake videos (at or above the accuracy threshold) \\ \hline
HENG & High engagement \\ \hline
LENG & Low engagement \\ \hline
ENG & Total engagement (HENG $+$ LENG) \\ \hline
DIS & Distraction \\ \hline
SO & Sleep onset \\ \hline
PSD & Power spectral density \\ \hline
AOI & Area of interest (eye tracking) \\ \hline
FPR & False positive rate (manipulated clip accepted as authentic) \\ \hline
FNR & False negative rate (genuine clip rejected as manipulated) \\ \hline
\end{tabular}
\end{table}

\subsection{Overview of Electroencephalography}
\label{sec:EEG}

EEG records electrical brain activity by measuring voltage differences between scalp electrodes. Originally a clinical tool for sleep and neurological disorders, it is now widely used in neuroscience to capture millisecond-scale neural responses. Electrode placement follows the international 10–20 system, with sites labeled by region: Fp (frontopolar), F (frontal), T (temporal), O (occipital), C (central), and P (parietal). The electrode layout used by our recording device is reproduced in the supplementary material for reference.

\textbf{The B-Alert X10} (10-channel EEG, 256~Hz) \cite{BAlertX10} is widely used in cognitive EEG research and avoids the impedance and motion artifacts common to high-density caps in long viewing tasks. It samples the Fz, F3, F4, C3, Cz, C4, P3, POz, and P4 sites of the 10–20 system. Signals are amplified, digitized, and streamed via Bluetooth from a portable head unit to a host computer.

\textbf{The B-Alert Lab} software \cite{BIOPAC} computes per-interval probabilities for four cognitive states: high engagement (HENG, sustained focus and active processing), low engagement (LENG, lower attention level), distraction (DIS, attention diverted from the task), and sleep onset (SO, reduced responsiveness). Total engagement (ENG) combines HENG and LENG, while drowsiness combines DIS and SO. Together these provide a continuous index of attentional and cognitive state during a task \cite{berka2007,JOHNSON2011}.

\subsection{Overview of Eye Tracking}
\label{sec:eye-tracking}
Eye tracking is a method of recording and analyzing eye movements in order to gain insight into an individual's thoughts and behaviors \cite{hashem2017}. This technique is used in multiple spheres such as medical research, communications, psychology, etc. \cite{luca2009,eberz2015}. In this project, the \textit{Gazepoint GP3 HD} eye tracker—which features a high-resolution biometric data capture system including eye gaze data, pupil diameter, heart rate, galvanic skin response (GSR), and a self-reporting dial—was utilized. The device operates with a sampling rate of 150 Hz at a distance of 50–80 cm.

\subsection{Overview of Deepfakes} \label{sec:deepfake} Using artificial intelligence applications to manipulate the actors in videos, particularly physically (eyes, mouth, lips, and face), is referred to as deepfakes. This expression is used due to the implementation of deep learning algorithms. In other words, it is a technology that alters videos and pictures of people by manipulating them and allowing them to be used for different purposes. This technology, which is used primarily to create fake news, videos, or images, has become widespread in recent years due to the development of high technology. Although producing deepfake video, image, or voice may not produce very successful results at first, with the use of modern technology, extremely high-quality results begin to emerge. This can be seen in recent advances in the deepfake field \cite{insider}, where the quality of deepfakes has increased rapidly to extremely high levels. For example, it is possible to generate images and videos of people that do not even actually exist, just like in \textit{https://thispersondoesnotexist.com} \cite{notexist} and \textit{https://www.whichfaceisreal.com} \cite{faceisreal}. Moreover, with more research and development, the quality and accuracy of deepfakes will continue to improve.

Several open-source libraries, such as DeepFaceLab \cite{deepfacelab}, FaceApp \cite{faceapp}, FaceSwap \cite{faceswap}, and Face2Face \cite{face2face}, have been developed to enable their creation, using AI components such as GANs, CNNs, and autoencoders.

\subsection{Threat Model}
\label{sec:threatmodel}
Attackers possessing advanced technical skills and extensive resources pose a significant threat with the creation of deepfakes. They are capable of producing convincing, high-quality deepfakes that are often indistinguishable from authentic videos. These attackers frequently leverage large, publicly available datasets to train deep learning models, resulting in outputs with high degrees of realism and precision.

There have been several notable deepfake incidents that illustrate this threat. For example, deepfakes of political leaders---such as fabricated videos of former U.S. President Barack Obama delivering misleading statements~\cite{mack2018obama}---have raised concerns about the potential for political misinformation. Similarly, a deepfake video of Facebook CEO Mark Zuckerberg appeared online in 2019~\cite{cbs2019zuckerberg}, showcasing how such content can challenge perceptions of authenticity on social media platforms. In the commercial sphere, a 2019 incident involved fraudsters using deepfake voice technology to impersonate the chief executive of a German parent company, inducing the head of its UK subsidiary to transfer \$243,000~\cite{stupp2019voice}.

\textbf{Operationalizing the evaluation:} In our lab setting, participants were exposed to crafted deepfake clips presented as authentic footage. Within the DECEIVE framework, attack success was quantified along two dimensions: (i) behavioral compromise—false acceptance of manipulated content, and (ii) subconscious evasion—absence of differential EEG or eye-tracking markers between real and fake. Each outcome was stored for iterative analysis, allowing us to identify videos that consistently bypass defenses as “successful” payloads. This operationalization frames our study as both an empirical test of susceptibility and an instantiation of DECEIVE’s evaluation pipeline.

\textbf{Why two levels of evasion.} A reader may reasonably object that if a manipulation is rejected only unconsciously, the viewer takes no protective action and the attack nonetheless succeeds. We agree: for an unaided viewer, behavioral compromise alone determines whether an attack works, and we state this explicitly rather than implying otherwise. The second dimension addresses a separate question, namely whether the viewer's perceptual system registers information that distinguishes authentic from manipulated content even when the viewer is consciously deceived.

This distinction determines which defenses remain available. If a deepfake is accepted behaviorally but still elicits a differential neural or ocular response, that signal exists within the observer and could in principle be surfaced---by instrumenting the viewer, or by directing attention toward the regions that already produce the response. If no such response occurs, there is nothing latent to surface, and defenses that depend on reading the human perceptual system cannot succeed at any level of sensor quality. The first dimension therefore characterizes present-day attack success; the second characterizes residual detectability, and hence the prospects for any defense that relies on human perception.

The distinction is practically relevant because consumer EEG and eye-tracking hardware has become inexpensive enough for large-scale deployment, and proposals for gaze-aware or neuroadaptive protective interfaces rest on the assumption that human perception yields a usable signal. Our results indicate that, for the content examined here, it does not. This reinforces rather than weakens the objection above: the attack succeeds against both conscious judgment and measurable perceptual processing, which is our principal reason for arguing that mitigation must come from automated detection, provenance and watermarking standards, and platform-level controls rather than from the viewer.

\section{Related Work}
\label{sec:relatedwork} 
Our study builds on a growing body of interdisciplinary research exploring deepfake detection using behavioral, neural, and physiological signals. Several prior studies suggest that neuro-physiological data may reveal implicit cues to manipulation, even when participants fail to consciously detect deception. For instance, Moshel et al. \cite{moshel2022real} found that EEG signals could reliably differentiate between real and AI-generated facial images, despite participants often misclassify them behaviorally. Similarly, Gupta et al. \cite{gupta2020eyes} introduced FakeET, an eye-tracking and EEG \emph{database} assembled to support AI-based deepfake detection; their EEG-based classifier reached an AUC of 0.55. Early work by Neupane et al. \cite{neupane2014neural,neupane2015multimodal,neupane2017neural} further suggested that the brain may subconsciously respond to synthetic media. However, more recent findings by Neupane et al. \cite{neupane2019crux} and Arisoy et al. \cite{arisoy2022human} challenge this view, reporting minimal neural distinctions—consistent with our own findings. While much of this research frames deepfake detection as a passive recognition task, our work evaluates these technologies and human perception within an active attack–defense scenario. While most prior work focuses on adversarial attacks against deepfake detectors—such as perturbing fake videos to evade machine classifiers \cite{hussain2020adversarial,neekhara2021adversarial}—our study differs by evaluating deepfake effectiveness as an attack vector against human perceptual and neuro-physiological defenses in a controlled laboratory setting.

In contrast to prior work, our study found no significant EEG or eye-tracking differences between real and deepfake videos, despite participants achieving high behavioral accuracy (~80\%). This suggests that under conditions involving dynamic video stimuli and familiar identities, neural signals may not always reflect perceptual discrimination. Our methodology expands the literature by introducing familiarity-based categories (FAM, BFAM, ALIKE), allowing us to examine how prior exposure influences both gaze behavior and susceptibility to deception.

\textbf{Differentiation from FakeET.} Because Gupta et al.~\cite{gupta2020eyes} study a superficially similar question, we state the differences explicitly. The two studies have different goals: they collect gaze and EEG as implicit labels to improve a CNN-based detector, evaluating success through gains in model performance, whereas we study the viewer as the target of the attack and ask whether deepfakes overcome human perceptual defenses. The stimuli also differ. FakeET draws on the Google/Jigsaw dataset, in which the depicted individuals are paid actors unfamiliar to participants; our stimuli show widely recognized public figures, so prior familiarity becomes a factor we manipulate rather than a constant we cannot vary. We further include a look-alike condition consisting of real, unmanipulated videos of people who resemble the target identity.

The look-alike condition serves two purposes. First, it provides a positive control for our instrumentation: our EEG measures distinguished look-alike from familiar and briefly familiar conditions ($p=0.017$, $0.04$, and $0.025$) and distinguished video trials from rest, while our gaze measures resolved significant differences across areas of interest within videos ($p<0.001$). Our instruments therefore detected differences where differences were present, so the absence of a real-versus-deepfake effect is unlikely to reflect measurement failure. Second, it separates identity mismatch from synthetic manipulation, since a look-alike presents the former without the latter. Participants identified human impostors with 95.24\% accuracy but struggled with synthetic ones, producing a 31.94\% false positive rate for familiar identities. FakeET includes neither a familiarity manipulation nor a comparable look-alike condition.

\begin{figure*}[hbt!]
  \centering
  \includegraphics[width=0.95\textwidth]{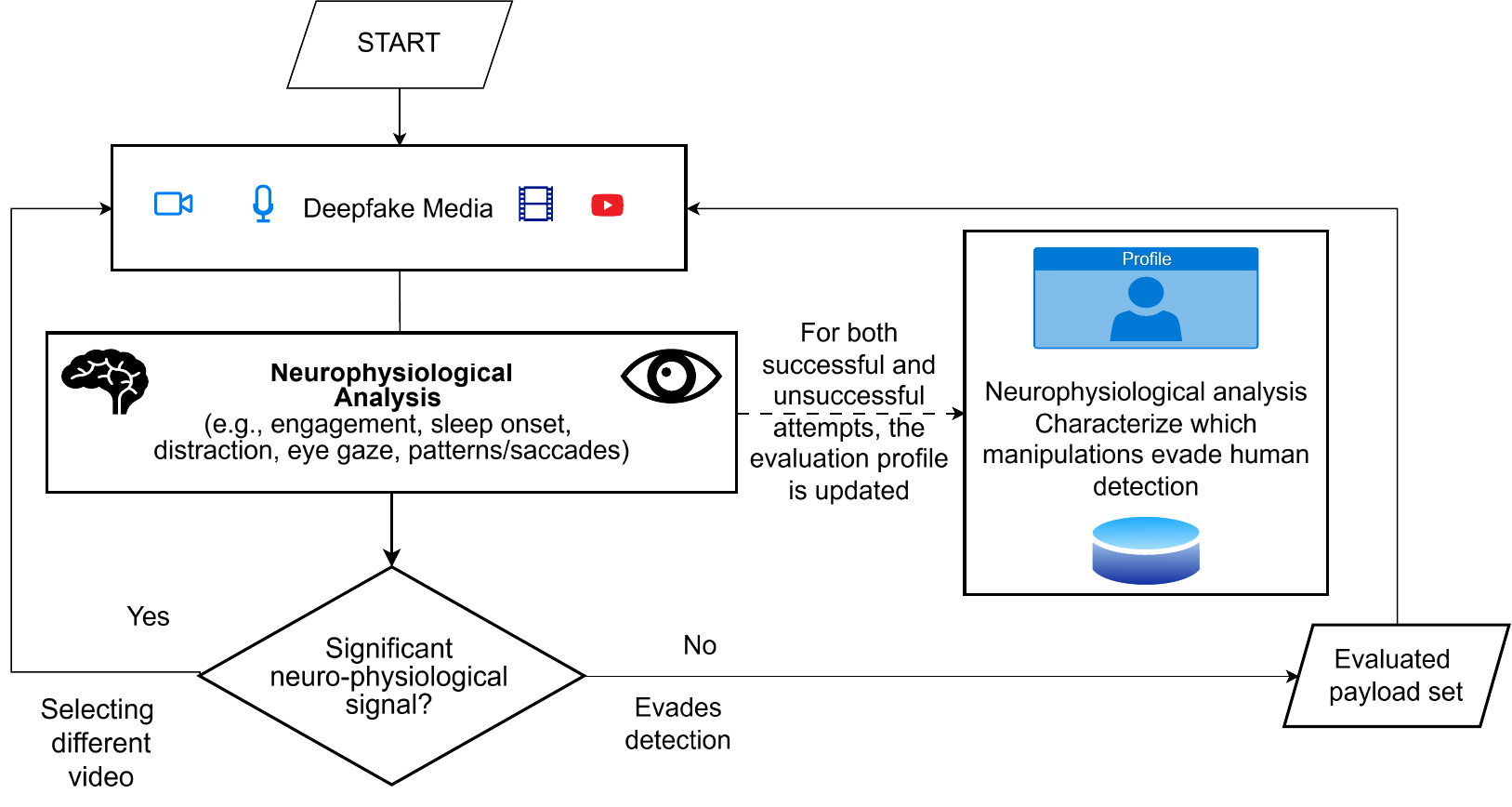}
  \caption{DECEIVE framework illustrating the primary stages of the cognitive–behavioral evaluation pipeline. The screening stage collects both the viewer's behavioral judgment and the neuro-physiological response; a payload advances only when undetected on both.}
  \label{fig:attack-pipeline}
\end{figure*}

Recent perceptual and EEG studies further inform our work: gaze fixations and perceptual accuracy vary with whether a video is real or face-swapped and with the generation method \cite{woehler2021perceptual,bozkir2024}; gaze-constrained generation reduces detectable artifacts \cite{wilson2023gazeconstraints,wilson2024uncanniness}; and EEG work links familiarity and emotional processing to deepfake perception \cite{tarchi2023eeg,tauscher2021eegbased,chakraborty2023survey,eiserbeck2023}. Our contribution combines EEG, eye-tracking, and behavioral data in a familiarity-aware framework, showing that neural signals alone may not distinguish real from fake in naturalistic contexts.

\section{The DECEIVE Framework}
\label{sec:deceive}
To formally model deepfake deception, we introduce the \textbf{DECEIVE} (Deepfake Exploitation of Cognitive Engagement and Implicit Visual Evaluation) framework. As illustrated in Figure~\ref{fig:attack-pipeline}, DECEIVE is a modality- and dataset-agnostic pipeline in which candidate deepfakes are repeatedly exposed to observers, and effectiveness is assessed along two dimensions: \textit{behavioral compromise} (misclassifying manipulated content as authentic) and \textit{subconscious evasion} (the failure of implicit signals---EEG, eye-tracking, or other biometric markers---to differentiate manipulated from authentic content). Content that bypasses both conscious and subconscious defenses is deemed a successful payload. Figure~\ref{fig:attack-pipeline} depicts the neuro-physiological screening stage explicitly; the behavioral judgment is collected at the same point in the pipeline, and a payload proceeds only if it is undetected on both dimensions. Each screened payload contributes to an \emph{evaluation profile}: a record, accumulated across both detected and undetected clips, of which manipulation characteristics coincide with successful evasion. It is this accumulated record, rather than any individual verdict, that indicates which perceptual vulnerabilities generalize across payloads. Beyond validating which deepfakes deceive viewers, DECEIVE gives defenders a structure to anticipate and mitigate such attacks. We instantiate it next through a controlled EEG and eye-tracking study quantifying how far state-of-the-art deepfakes evade both conscious and subconscious detection.

\section{Concrete DECEIVE Instantiation: Our Study Design \& Procedures}
\label{sec:design}
\subsection{Ethical Considerations and Data Privacy}
\label{sec:ethicalandprivacy}
This study was approved by the university's Institutional Review Board. Participants were recruited via the bulk email request from the university and were compensated with a \$30 incentive for their involvement, approximately 1.5 hours. While the entire session lasted ~90 minutes, including calibration tasks, the actual time spent watching videos averaged \textbf{25–30 minutes}. Participants were informed that they could discontinue the experiment at any time without explanation. EEG and gaze data were recorded anonymously, and strict measures were taken to preserve the confidentiality of all data collected during the study. Participants were fully debriefed at the end of the experiment and informed that the study involved the use of deepfake videos, in accordance with our IRB-approved protocol. We recognize the dual-use character of this work. Only publicly available videos and established datasets were used, no harmful or defamatory material was included, and we release only curated, non-harmful stimuli. The purpose of quantifying perceptual vulnerabilities is defensive: to inform detection tooling, provenance mechanisms, and user-facing protections, as set out in Section~\ref{sec:threatmodel}. Further ethical motivations and risk-mitigation detail are documented in the supplementary material.

\subsection{Experiment Design}
\label{sec:experimentdesign}
Participants acted as defenders, viewing a naturalistic mix of authentic and manipulated (deepfake and impersonation) videos without being told in advance which was which until the post-experiment debrief. Each trial functioned as a candidate payload (Figure~\ref{fig:attack-pipeline}): a clip detected consciously or via differential EEG/eye-tracking is rejected, while one undetected at both levels is scored as a successful evasion.

We did not employ event-related potential (ERP) analysis. ERP does not require stimuli of equal duration; it requires precise synchronization with a defined stimulus onset, and total clip length is immaterial beyond the analysis window. Our protocol provides no such onset marker: video presentation was self-paced, no fixation cross or synchronized event marker preceded each trial, and our cognitive-state metrics are computed over one-second epochs. Our design instead targets sustained cognitive engagement during naturalistic viewing, for which epoch-level measures are appropriate. We randomized video presentation to avoid order effects and to assess how viewing time shaped cognitive and behavioral responses.  

To address fatigue and stress, clips lasted approximately 13 seconds on average, reflecting the length of most online deepfake scams and aligning with user habits (as supported by response times). Each participant viewed 104 videos, ensuring variability and sufficient data while minimizing burden, consistent with prior work (e.g.,~\cite{gupta2020eyes}). We further introduced short breaks and visual cues (e.g., progress bars) to reduce fatigue and maintain engagement.

\begin{figure*}[!ht]	
		\centering	
		\includegraphics[width=1\textwidth]{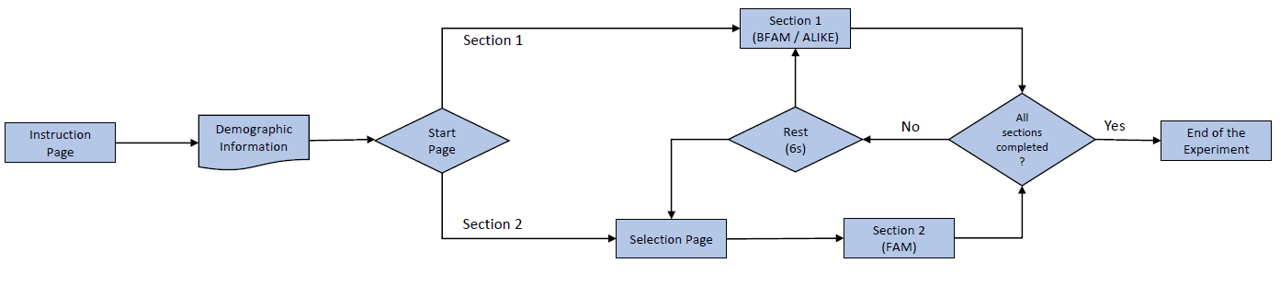}
		\caption{The flow diagram representing the design of the experiment}
		\label{fig:flowchart}
\end{figure*} 

\subsubsection {\textbf{Dataset}}
\label{dataset}
We combined the public \textit{Celeb-DF} dataset \cite{Celeb-DF} with a curated YouTube set (\textit{COT-DB}); each participant viewed 104 short ($\sim$13\,s) clips (52 real, 52 manipulated) across three familiarity-based conditions (full counts, celebrity lists, and curation details in the supplementary \emph{Design of the Experiment}). \textbf{Familiar (FAM):} highly recognizable individuals (17 Celeb-DF celebrities plus Cruise, Obama, Trump); participants also self-selected three further familiar celebrities. \textbf{Briefly Familiar (BFAM):} three unnamed Celeb-DF actors shown anonymously (Person~1--3). \textbf{Look-alike (ALIKE):} authentic videos of real people who physically resemble Cruise, Obama, or Trump (genuine doppelg\"angers, not manipulations), used as a baseline against which deepfakes are compared. Our selection strategy ensured diverse manipulations, actors, and realistic scenarios, while personalized video choices further enhanced experimental realism.

\subsubsection {\textbf{Task Design}}
\label{sec:taskdesing}
For our experiment, we developed a custom software application comprising eight sections. The \textit{Instructions Page} provided task details, and the \textit{Demographic Information Page} collected participant data. In \textit{Section~1}, participants viewed BFAM and ALIKE videos, while \textit{Section~2} began with the \textit{Selection Page}, where they chose three celebrities before being shown all FAM videos. The \textit{Experiment End Page} redirected participants to the post-questionnaire. Two 6-second \textit{Rest Pages} were included to reduce fatigue and refocus attention between sections. Each stage featured a Progress Bar to provide feedback and sustain engagement. An overview of the experimental flow is presented in Figure~\ref{fig:flowchart}, and the supplementary material illustrates the setup with the eye tracker and EEG headset.

\textbf{Participant Procedure and Task Flow:} Participants completed the experiment on a designated computer. After reviewing on-screen instructions and logging demographic data on the \textit{Demographic Information Page}, they proceeded to the \textit{Experiment Start Page}, which randomly directed them to Section~1 or 2 (all participants completed both). Each actor was introduced via a short video, after which participants viewed randomized sequences of real and fake videos. They were instructed to attend carefully to each video, as it represented the actual subject under study, and then press “Continue” to proceed.

\textbf{Study structure.} The experiment had two sections (per-section and per-celebrity counts in the supplementary \emph{Design of the Experiment}). Each clip could be replayed. After each video participants answered two yes/no questions: (1) whether the person shown in the (silent) clip was the identity introduced before the block (for example, ``Do you think the speaker in this snippet video is Barack Obama?''), and (2) whether they had seen the clip before. Participants were never asked to classify a clip as ``real'' or ``fake''. A ``Yes'' response is scored as acceptance of the clip as an authentic depiction of the named identity, and a ``No'' response as rejection; all reported rates derive from this mapping. Participants were not told in advance whether any clip was authentic or manipulated. EEG and eye-tracking captured implicit responses alongside the explicit judgments; all procedures were IRB-approved with informed consent and a debrief. %
\subsection{Experimental Protocol}
\label{sec:experimentalprotocol}
\subsubsection{\textbf{Participant Recruitment and Preparation}}
\label{sec:demographics}

There is no universally defined minimum sample size for EEG studies, as optimal sample size depends on the specific research design and intended analyses. However, prior literature suggests that a sample of 20–30 participants is generally sufficient to achieve acceptable statistical power in cognitive neuroscience research \cite{Dupont2010,Machin1999,Altman1991}. 

We recruited 25 healthy participants (13 female, 12 male) with no history of photosensitive epilepsy or implanted devices. One was excluded due to eye-tracking calibration failure. All participants gave informed consent. Participants were not told that the study focused on deepfakes, in order to reduce bias; the per-clip judgment they provided is the identity question described in Section~\ref{sec:taskdesing}. Post hoc EEG quality checks excluded four participants for noisy signals, leaving 20 (10 female, 10 male) in the final EEG analysis; eye-tracking and behavioral data from 24 participants (12 female, 12 male) were retained. Participants were recruited via the university's central mailing list and volunteer research pool, and included 9 undergraduates, 7 master's students, and 8 Ph.D. candidates spanning ages 18–55 (8 in 18–24, 8 in 25–30, 4 in 31–40, 4 in 41–55). This sourcing ensured independent participation and avoided shared bias from peer referral; it also reflects standard practice in EEG-based human-subject research \cite{berka2007,gupta2020eyes}, where in-person, lab-based protocols preclude social-chain recruitment. A full demographic breakdown is additionally tabulated in the supplementary material.

A post-hoc power analysis (G*Power 3.1; paired design, $\alpha=0.05$, $d=0.5$) indicated $\sim$20 participants suffice for 80\% power, matching our EEG sample (n$=$20; behavioral/eye-tracking n$=$24); a Bayesian estimate gave a credible moderate effect (mean 0.35, 94\% HDI [0.02, 0.65]). Post-hoc power is supportive context, not confirmatory. Comparable sizes appear in related neuro-security work \cite{neupane2014neural,neupane2015multimodal,neupane2017neural,neupane2019crux,gupta2020eyes}. %
\subsection{Data Collection}
\label{sec:datacollection}
After fitting the B-Alert headset and seating the participant, we ran the standard 15-minute B-Alert calibration (three 5-minute tasks: 3-Choice Vigilance, Eyes-Open, Eyes-Closed) to build participant-specific profiles for the cognitive-state metrics (HENG, LENG, DIS, SO). Task details and reference thresholds are in the supplementary material. In parallel with EEG preparation, the Gazepoint device \cite{gazepointgp3} was positioned 65~cm from the user and calibrated using Gazepoint's standard 5- or 9-point Calibration Task. Once both EEG and eye-tracking were set up, participants began the experiment by launching our custom Windows Forms application. The supplementary material shows the eye tracker and EEG headset positioned on a participant. Participants then completed the Post-Test Questionnaire (see the supplementary material) via the link on the \textit{Experiment End Page}.

\subsection{Post Experiment Survey}
After completing the experiment, we conducted a survey to gather insights into the participants' awareness and understanding of deepfakes. Specifically, participants were asked about their prior knowledge of the term 'deepfake,' their ability to distinguish deepfakes from authentic content, and their overall relationship with technology such as computers and smartphones. The results of this survey will be detailed in the Results (Section \ref{sec:result}). The survey aimed to assess participants' awareness of deepfakes and their confidence in identifying them, to explore their online video-watching habits, particularly on online social platforms, and to learn participants' beliefs and attitudes regarding deepfake detection and its implications.

\section{Data Processing and Analysis Methods}
\label{sec:analysis}
\subsection{Brainwave Data}
\label{neuraldataanalysis}
Cognitive-state metrics are derived from one-second epochs via a four-class quadratic discriminant analysis on 1\,Hz EEG power spectral densities at FzPOz, CzPOz, FzC3, C3C4, and F3Cz \cite{biopac2}; the B-Alert Lab software \cite{BIOPAC} outputs per-epoch probabilities for HENG, LENG, DIS, and SO, with the highest-probability class assigned.

EEG data were collected from five channels (FzPO, CzPO, FzC3, C3C4, F3Cz). Power Spectral Density (PSD) was computed using Fast Fourier Transform (FFT) across 1–40 Hz, with emphasis on theta (3–7 Hz), alpha (8–13 Hz), beta (13–29 Hz), and gamma (30–40 Hz) bands. High- and low-pass filters were applied for signal refinement. PSD characterizes neural power distribution associated with cognitive states; in the context of deepfake detection, band-specific modulations may indicate cognitive effort in identifying anomalies, whereas the absence of such differences suggests successful neural evasion by deepfakes.

\subsection{Eye Tracking Data}
\begin{figure}[!htb]
    \centering         
    \includegraphics[width =0.8 \linewidth]{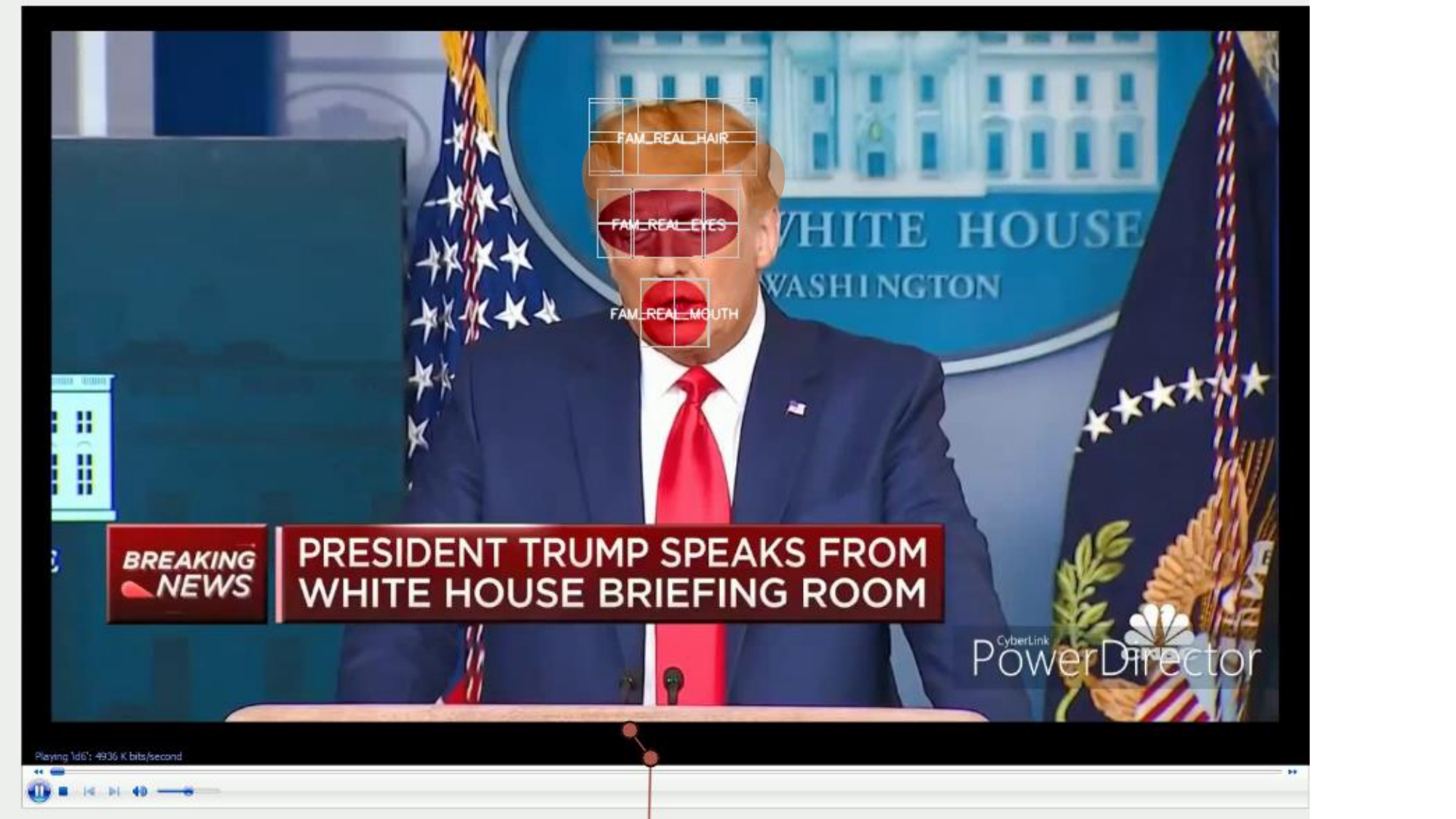}
    \caption{Gaze plot showing the area of interest (AOI) marked}
    \label{fig:aoi-marked}
\end{figure}

Using the Gazepoint GP3 HD device \cite{gazepointgp3}, we recorded gaze metrics including fixations, gaze duration, and saccades. Fixations and gaze duration capture the time the eye remains relatively stationary on a region, while saccade magnitude and direction indicate the distance and angle between fixations. Areas of interest (AOIs) were defined for each video, focusing on the actors’ eyes, mouth, and head. Data on gaze time, fixations, and revisit counts were collected using Gazepoint Analysis Professional Edition. AOI regions are illustrated in Figure~\ref{fig:aoi-marked}.

\subsection{Statistical Testing Methodology}
\label{statisticalmethod}
We used IBM SPSS Statistical Software \cite{IBMSPSS} for analysis, evaluating participant responses across false–true classifications, video types, and real–fake perspectives. All results are reported at a significance level of $\alpha=0.05$. Initial differences across video epochs were assessed using the Friedman test; where significance was detected, pairwise comparisons were conducted with the Wilcoxon Signed-Rank Test (WSRT) \cite{wilcoxon}. Effect sizes were computed as $r = Z/\sqrt{N}$, where $Z$ is the test statistic and $N$ the number of observations, and interpreted according to Cohen’s criteria \cite{cohen1977statistical}: $r > 0.1$ (small), $>0.3$ (medium), $>0.5$ (large). Bonferroni corrections \cite{Bonferroni} were applied to all significant pairwise comparisons.

\section{Results}
\label{sec:result}
\subsection{Task Performance Results}
\label{sec:task}
Participants judged whether the speaker shown was the named identity. We treat a ``Yes'' response as acceptance of the clip as an authentic depiction and a ``No'' response as rejection, and we take \emph{authentic} as the positive class throughout. This yields four outcomes: \emph{Real as Real}, a genuine clip accepted (true positive); \emph{Fake as Real}, a manipulated clip accepted (false positive); \emph{Fake as Fake}, a manipulated clip rejected (true negative); and \emph{Real as Fake}, a genuine clip rejected (false negative). The false positive rate (FPR) is the proportion of manipulated clips accepted, and the false negative rate (FNR) the proportion of genuine clips rejected. In security terms, the false acceptance rate is identical to FPR and the true rejection rate is its complement; we note this correspondence once and use FPR and FNR consistently in the remainder of the paper.

Across 24 participants, overall accuracy was 80.77\% (recall 0.88, precision 0.77, F$=$0.82). The mean response time was 9795\,ms and was significantly longer for incorrect than correct judgments (11{,}684 vs.\ 9345\,ms; $p<.001$, WSRT), suggesting hesitation on harder decisions. Category breakdowns appear in Tables~\ref{tab:overall-task} and~\ref{tab:category-task}.

\begin{table}[ht]
\centering
\caption{Overall task performance: participants showed $>$25\% FPR, indicating susceptibility to deepfake attacks.}
\label{tab:overall-task}
\resizebox{\columnwidth}{!}{
\begin{tabular}{|l|r|r|r|r|r|r|}
\hline
\textbf{Overall} &
  \textbf{Avg RT (ms)} &
  \textbf{Accuracy (\%)} &
  \textbf{Recall} &
  \textbf{Precision} &
  \textbf{F-measure} &
  \textbf{Accuracy (\%)} \\ \hline
\textbf{(n=24)} & 9795 &  &  &  &  &  \\ \hline

Fake as Fake &
  9159 &
  \textbf{73.32} &
  \multirow{4}{*}{0.88} &
  \multirow{4}{*}{0.77} &
  \multirow{4}{*}{0.82} &
  \multirow{4}{*}{80.77} \\ \cline{1-3}

Fake as Real (FPR) &
  11435 &
  \cellcolor[gray]{0.8}\textbf{26.68} &
  & & & \\ \cline{1-3}

Real as Fake &
  12247 &
  11.78 &
  & & & \\ \cline{1-3}

Real as Real &
  9499 &
  88.22 &
  & & & \\ \hline  
\end{tabular}
}
\end{table}

\subsubsection{\textbf{Category-Based Results}}
Analyzing the false positive rate---manipulated clips accepted as authentic---we found an overall rate of 26.68\%, with substantial variation across categories (Table~\ref{tab:category-task}).

\begin{table}[ht]
\footnotesize
\centering
\caption{Task performance by video category. The positive class is an authentic depiction of the named identity, so \emph{Fake as Real} is the false positive rate.}
\label{tab:category-task}
\resizebox{\columnwidth}{!}{
\begin{tabular}{|l|l|l|l|l|l|}
\hline
\textbf{OVERALL (n=24)} & \textbf{\%} & \textbf{recall}                & \textbf{precision}             & \textbf{f-measure}             & \textbf{accuracy (\%)}          \\ \hline \hline
\textbf{ALIKE}          &    & \multirow{5}{*}{0.98} & \multirow{5}{*}{0.93} & \multirow{5}{*}{0.95} & \multirow{5}{*}{95.24} \\ \cline{1-2}
Fake as Fake & \textbf{92.26}  &  &  &  &  \\ \cline{1-2}
Fake as Real & \textbf{7.74} \cellcolor[gray]{0.8}   &  &  &  &  \\ \cline{1-2}
Real as Fake & 1.79  &  &  &  &  \\ \cline{1-2}
Real as Real & 98.21 &  &  &  &  \\ \hline \hline
\textbf{FAM}            &    & \multirow{5}{*}{0.90} & \multirow{5}{*}{0.74} & \multirow{5}{*}{0.81} & \multirow{5}{*}{79.22} \\ \cline{1-2}
Fake as Fake & \textbf{68.06} &  &  &  &  \\ \cline{1-2}
Fake as Real & \textbf{31.94} \cellcolor[gray]{0.8}  &  &  &  &  \\ \cline{1-2}
Real as Fake & 9.61  &  &  &  &  \\ \cline{1-2}
Real as Real & 90.39 &  &  &  &  \\ \hline \hline
\textbf{BFAM}         &    & \multirow{5}{*}{0.72} & \multirow{5}{*}{0.78} & \multirow{5}{*}{0.75} & \multirow{5}{*}{75.69} \\ \cline{1-2}
Fake as Fake & \textbf{79.63} &  &  &  &  \\ \cline{1-2}
Fake as Real & \textbf{20.37} \cellcolor[gray]{0.8} &  &  &  &  \\ \cline{1-2}
Real as Fake & 28.24 &  &  &  &  \\ \cline{1-2}
Real as Real & 71.76 &  &  &  &  \\ \hline \hline
\textbf{FAM + BFAM}    &    & \multirow{5}{*}{0.87} & \multirow{5}{*}{0.75} & \multirow{5}{*}{0.80} & \multirow{5}{*}{78.52} \\ \cline{1-2}
Fake as Fake & \textbf{70.37} &  &  &  &  \\ \cline{1-2}
Fake as Real & \textbf{29.63} \cellcolor[gray]{0.8} &  &  &  &  \\ \cline{1-2}
Real as Fake & 13.33 &  &  &  &  \\ \cline{1-2}
Real as Real & 86.67 &  &  &  &  \\ \hline
\end{tabular}
}
\end{table}
 FPR peaked in the FAM category at 31.94\%, indicating high misidentification, while BFAM and ALIKE were lower at 20.37\% and 7.74\%, respectively. Combined, FAM and BFAM yielded an FPR of 29.63\%. Overall participant accuracy was 80.77\%, but dropped to 73.32\% on fake videos. Precision also varied, with FAM showing the lowest at 74\%. Yet, combined FAM and BFAM accuracy reached 78.52\%, underscoring the difficulty in distinguishing real from deepfakes. Turning to the false negative rate---genuine clips rejected as manipulated---the FAM category showed a low 9.61\%, suggesting participants were generally reliable at recognizing genuine videos despite elevated FPR and reduced precision.

\subsubsection{\textbf{Easy vs Difficult Deepfake Videos}}
Task performance analysis showed large variation in FAM videos. Within the curated COT-DB subset, average fake-video accuracy was 46.29\%, so clips below this threshold were labeled ‘difficult’ (D-Fake) and the rest ‘easy’ (E-Fake). D-Fakes took longer to identify (13,671 ms vs. 9,983 ms) and had lower overall accuracy (54.17\% vs. 81.71\%), reflecting the greater challenge of detecting higher-quality fakes. These results are summarized in the supplementary material.

\subsubsection{\textbf{Post Experiment Results}}
As noted earlier, participants answered two questions after each video. Most videos (97.40\%) were reported as previously unseen, \& excluding the few familiar ones did not alter results. A post-experiment questionnaire (see the supplementary material) provided further insights. Demographically, 45.8\% of participants identified as Asian, 29.2\% as White, 16.7\% as Black/African American, and the remainder as Hispanic. Most had heard of deepfakes within the past five years. While 37.5\% believed they could detect deepfakes in daily life, 75\% reported doing so during the experiment, with 8\% expressing confidence in this ability. Finally, 71\% perceived deepfakes as a personal threat, and 83\% as a national/international security threat.

\subsection{Brainwave Results}
\label{sec:neuralresult}
\subsubsection{\textbf{Video Trials vs Rest Trial}}
Engagement was significantly higher during the video sections than during rest. Comparing average percentage-of-frequency (pfr\_) metrics---pfrLENG, pfrHENG, pfrENG, pfrDIS, and pfrSO---between the six-second rest period separating \textit{Section 1} and \textit{Section 2} and the FAM, BFAM, and ALIKE sections, HENG differed significantly by the Wilcoxon Signed-Rank Test (WSRT), indicating that participants were more engaged while viewing than at rest.

\subsubsection{\textbf{Video Category Results}}
Comparing average pfr metrics (detailed in the supplementary material) with the WSRT, we found \emph{no} significant fake-vs-real difference within any category. Significant differences emerged only between ALIKE and the FAM/BFAM categories (e.g., LENG/HENG for real ALIKE vs.\ BFAM, $p=0.017/0.04$; ENG for fake ALIKE vs.\ FAM, $p=0.025$; see the supplementary material), indicating greater engagement for FAM and BFAM than ALIKE videos.

\subsubsection{\textbf{Raw Data Results}}
\label{sec:rawdataanalysis}

We analyzed B-Alert EEG data (alpha, beta, theta, 1–40 Hz) across fake and real videos. Some beta variations appeared with low-pass filtering, but these were not consistently supported.
Consequently, as detailed in the supplementary material, our analysis did not reveal any substantial differences between the EEG signals elicited by fake and real videos.

Notable differences emerged in cognitive analysis between ALIKE vs. FAM and ALIKE vs. BFAM conditions. Raw data indicated higher spectral power density (PSD) for the FAM group relative to the others. Specifically, the alpha signal of FAM\_FAKE exceeded 40 dB, whereas BFAM\_FAKE and ALIKE\_FAKE fell below –40 dB. Likewise, FAM\_REAL ranged between 20–40 dB, while BFAM\_REAL and ALIKE\_REAL ranged from –60 to –50 dB. These alpha-band disparities highlight clear frequency-component differences across video types, which remained robust after applying high- and low-pass filters. Detailed spectra are provided in the supplementary material.

\subsection{Eye Tracking Results}
\label{sec:eyetracking}

We analyzed participants’ eye-tracking data to examine fixation time differences across classifications. As detailed in the supplementary material, no significant differences appeared when videos were correctly identified as real or fake. However, fixation times varied for misclassifications: 'Fake as Fake' responses were longer than 'Fake as Real' ($p=0.003$), and 'Real as Fake' longer than 'Real as Real' ($p=0.007$). These results suggest participants devoted more visual attention to videos they believed were fake, regardless of ground truth, indicating greater scrutiny when content was seen as untrustworthy.

We extended our analysis to eye-tracking data (detailed in the supplementary material), examining average metrics across FAM, BFAM, and ALIKE videos. Comparisons between fake and real videos revealed no statistically significant differences in fixation duration, saccade magnitude, or direction. Participants’ visual behaviors were thus similar across categories, underscoring the difficulty of distinguishing real from deepfakes using eye-tracking alone. Representative gaze plots by category are provided in the supplementary material.

We analyzed Areas of Interest (AOIs), eyes, mouth, hair, progress bar, and button locations (Figure~\ref{fig:aoi-marked}) to assess gaze allocation. No significant differences were found between real and fake videos. Across conditions, eyes and mouth received significantly more fixations and revisits than hair ($p<0.001$), confirming them as primary focal regions. Hair in fake videos drew comparatively longer viewing times and revisits, likely reflecting subtle artifacts, though not statistically significant. Progress bar fixations were slightly lower than other AOIs, suggesting participants maintained attention on the videos. Representative gaze distributions appear in the supplementary material.

\section{Discussion}
\label{sec:discussion}
Our study examined behavioral and neuro-physiological responses to deepfake versus real videos. Using the Celeb-DF and COT-DB datasets (97.4\% of clips were new to participants, reducing prior-exposure bias), we found no reliable distinction between real and fake videos behaviorally, neuro-physiologically, or in eye-tracking. Response times were longer for D-Fake than E-Fake videos, suggesting viewing time proxies certainty of authenticity and that higher-quality fakes are harder to detect. With no differential neural or ocular signal to draw on, viewers had no implicit cue---confirming successful subconscious evasion, in which deepfakes bypass both conscious and subconscious defenses.

Participants struggled most with familiar individuals: FAM D-Fake videos yielded a 32\% FPR and only 23.15\% accuracy. Familiarity thus offered no protection participants were worse at spotting fakes of well-known celebrities than lesser-known ones, though better at recognizing genuine videos. We attribute this to familiarity-induced overconfidence, the high fidelity of modern fakes, an authenticity bias toward recognizable faces, and weaker contextual cues for celebrities, underscoring the growing risk as deepfake quality advances.

An alternative explanation deserves consideration: generators may have been trained on more footage of widely recognized individuals, so that FAM deepfakes are technically better rather than FAM viewers being more susceptible. Our design addresses this in part. The Celeb-DF component of our stimuli holds generator exposure approximately constant: all 59 Celeb-DF identities are celebrities sampled from YouTube interviews, the manipulations are produced by swapping faces across every pair of those subjects using a single synthesis pipeline, and our FAM and BFAM conditions both draw from this pool. Familiarity, not generator exposure, is what we varied between them, and the false positive rate nonetheless differed substantially (31.94\% vs.\ 20.37\%). The ALIKE condition points the same way: those clips are authentic and unmanipulated, so synthesis quality cannot account for the 95.24\% accuracy observed there. We do not, however, regard the alternative as excluded. Our curated COT-DB material depicts exceptionally heavily filmed public figures and was sourced externally rather than generated under a controlled pipeline, so prominence and generation quality are confounded within it; the D-Fake threshold was computed on this subset, meaning the clips that produced our largest familiarity effect are also those for which the training-exposure account is most plausible. We did not measure per-clip technical fidelity. Separating viewer familiarity from generator training exposure would require holding identity prominence fixed while varying the volume of available source footage, which we leave to future work.

Two features of the BFAM condition further qualify this comparison. BFAM comprises only three identities, so the conditions are not matched in breadth. More importantly, participants shifted their response criterion toward rejection for unfamiliar faces: genuine BFAM clips were rejected at 28.24\%, against 9.61\% for FAM. Part of the lower BFAM false positive rate therefore reflects a greater willingness to answer ``no'' when the depicted person could not be recognized, rather than better discrimination as such---overall accuracy was in fact lower for BFAM (75.69\%) than for FAM (79.22\%). The familiarity effect we report is accordingly better characterized as a shift in the direction of errors than as a uniform loss of sensitivity.

Brainwave analysis showed significant rest-versus-trial differences, confirming the videos engaged participants; engagement was lowest for ALIKE videos, which were also easiest to detect, hinting at a link between engagement and discernment. Stable SO (sleep onset) and DIS (distraction) indicate no fatigue or stress. Eye-tracking showed consistent focus on eyes and mouth, with somewhat more attention to hair in fake videos---possibly reflecting artifacts, though hair was the least-gazed region and thus less reliable.

Raw-data and model analysis revealed distinct neural patterns by category, with FAM videos eliciting the highest engagement and 76.7\% classification accuracy, while BFAM and ALIKE were harder to separate. Consistent with prior work showing synthetic voices may be neurally indistinguishable from real ones \cite{neupane2019crux}, our results suggest that in real-world attacks combining manipulated audio and video, users may face no identifiable cues---and those with limited awareness, who often trust digital content without verification, are especially vulnerable.

\textbf{Why our neural results differ from FakeET.} Gupta et al.~\cite{gupta2020eyes} report an error-related negativity for fake videos, whereas we find no differential response. Several methodological differences plausibly account for this. Their effect is a transient ERP component at approximately 1.3\,s, while our measures aggregate cognitive state across the full clip and cannot resolve components at that timescale. Their gaze effect depends on viewers exploring a wider scene, whereas our head-and-shoulders footage offers little opportunity for such exploration. Finally, their participants performed an explicit two-alternative real-versus-fake forced choice, whereas ours made an identity judgment; the error-monitoring process that generates the ERN may therefore not have been engaged by our protocol. We consequently read our result as identifying the conditions under which implicit signals are available, rather than as contradicting theirs.

Although most participants (22/24) knew the term ``deepfake'' and were confident in detecting them, the absence of consistent neuro-physiological differentiation shows such confidence does not translate into subconscious discrimination---echoing prior findings on overconfidence in phishing detection and the need for objective detection over self-assessment.

\textbf{Study Limitations:} The laboratory setting may not capture real-world encounters, and findings are specific to the video types tested; prior knowledge of deepfakes may also have influenced performance. Our cognitive-state measures are computed over one-second epochs and are therefore insensitive to transient, time-locked components; an ERP-scale effect such as the error-related negativity Gupta et al.~\cite{gupta2020eyes} report at approximately 1.3\,s would not be resolvable by our analysis, so our null result concerns sustained differential engagement rather than transient responses. Our sample also lacked broad demographic diversity, precluding analysis of age, gender, cognitive ability, or technical familiarity, and our stimuli were silent, so the findings speak to visual deception alone and may understate susceptibility to multi-modal deepfakes that combine manipulated video with synthesized speech. Finally, our gaze analysis emphasized spatial AOIs; explicit temporal-feature analysis (e.g., scanpath modeling, saccade velocity) is left to future work.

\section{Conclusion}
\label{sec:conclusion} This work introduced DECEIVE, a framework for evaluating deepfake deception through both behavioral and neuro-physiological screening. By instantiating DECEIVE in a controlled laboratory study, we exposed participants to real, deepfake, and look-alike videos while measuring their explicit judgments and implicit EEG/eye-tracking responses.

Our findings validate the core premise of DECEIVE: high-quality deepfakes evade both conscious and subconscious defenses. Participants did not reliably distinguish real from manipulated content and showed no statistically significant neuro-physiological or ocular markers differentiating the two, demonstrating that such content functions as a successful payload; look-alike videos, by contrast, elicited distinct responses. The absence of subconscious differentiation also limits the value of neuro-physiological signals as detection features, suggesting deepfake deception is rooted in human perception itself.

In sum, DECEIVE operationalizes both a conceptual framework and an empirical evaluation method for deepfake deception. By demonstrating that humans lack reliable behavioral and neuro-physiological defenses against high-quality deepfakes, our work highlights the urgent need for robust automated countermeasures and situates DECEIVE as a foundation for future evaluation of synthetic media.

\begin{credits} \subsubsection{\ackname} We thank the anonymous reviewers and our shepherd for their invaluable comments, which substantially improved this paper. \end{credits}

\bibliographystyle{splncs04}
\bibliography{references}

\end{document}